\documentclass{article}

\usepackage{arxiv}
\usepackage{amsmath}
\usepackage[utf8]{inputenc} 
\usepackage[T1]{fontenc}    
\usepackage{hyperref}       
\usepackage{url}            
\usepackage{booktabs}       
\usepackage{amsfonts}       
\usepackage{nicefrac}       
\usepackage{microtype}      
\usepackage{cleveref}       
\usepackage{lipsum}         
\usepackage{graphicx}
\usepackage{natbib}
\usepackage{doi}
\usepackage{multirow}
\usepackage[table]{xcolor}

\newtheorem{lemma}{\textbf{Lemma} }
\newtheorem{proof}{\textbf{Proof} }
\newcommand{\argmin}[2]{\operatornamewithlimits{argmin}\limits_{#1}{\;\;#2}}

\newcommand{\mino}[2]{\operatornamewithlimits{min}\limits _{#1}{\;\;#2}}
\def \*#1{\mathbf{#1}}   
\def \r{\mathbf{r}}

\title{STAIC regularization for   spatio-temporal  image  reconstruction}

\date{}

\newif\ifuniqueAffiliation
\uniqueAffiliationtrue

\ifuniqueAffiliation 
\author{ Deepak G Skariah \\
	Department of Electrical Engineering \\
	IISc \\
	Bengaluru, 560012 \\
	\texttt{deepaks@iisc.ac.in} \\
	\And
	Muthuvel Arigovindan \\
	Department of Electrical Engineering\\
	IISc\\
		Bengaluru,560012\\
	\texttt{mvel@iisc.ac.in} \\
}
\else
\usepackage{authblk}

\newbox{\orcid}\sbox{\orcid}{\includegraphics[scale=0.06]{orcid.pdf}} 
\author[1]{%
	\href{https://orcid.org/0000-0000-0000-0000}{\usebox{\orcid}\hspace{1mm}David S.~Hippocampus\thanks{\texttt{hippo@cs.cranberry-lemon.edu}}}%
}
\author[1,2]{%
	\href{https://orcid.org/0000-0000-0000-0000}{\usebox{\orcid}\hspace{1mm}Elias D.~Striatum\thanks{\texttt{stariate@ee.mount-sheikh.edu}}}%
}
\affil[1]{Department of Electrical Engineering, IISc Bengaluru}
\affil[2]{Department of Electrical Engineering, IISc Bengaluru}
\fi

\renewcommand{\shorttitle}{\textit{STAIC} Algorithm}

\hypersetup{
pdftitle={A template for the arxiv style},
pdfsubject={q-bio.NC, q-bio.QM},
pdfauthor={David S.~Hippocampus, Elias D.~Striatum},
pdfkeywords={First keyword, Second keyword, More},
}

\begin{document}
\maketitle

\begin{abstract}
	We propose a regularization-based image restoration scheme for 2D  images recorded over time  (2D+t). We design an infimal convolution-based regularization function which we call
	spatio-temporal Adaptive Infimal Convolution (STAIC) regularization. We formulate the infimal convolution in the form of an additive decomposition of the 2D+t  image such that the
	extent of spatial and temporal smoothing is controlled in a spatially and temporally varying manner.  This makes the regularization adaptable to the local characteristics of the motion
	leading to an improved ability to handle noise.    We also develop a minimization method for image reconstruction by using the proposed form of regularization.  We demonstrate
	the effectiveness of the proposed regularization using TIRF images recorded over time and compare with some selected existing regularizations. 
\end{abstract}


 \section{Introduction}
 \label{sec:introduction}
 
 Image restoration is an inverse problem \cite{katsaggelos1989iterative} where a higher quality image estimate is generated from a corrupted observation by  exploiting   
 knowledge of image statistics.
 Regularization methods \cite{engl1996regularization} constitute an  important category among methods for biomedical image restoration \cite{rangayyan2004biomedical}. 
 Imaging modalities where  regularization methods have been successful in restoration of corrupted images
 include MRI \cite{ramani2012regularization,viswanath2020image},  Widefield Microscopy \cite{arigovindan2013image,li2018accurate} and Total Internal Reflection (TIRF) 
 Microscopy \cite{fan2019one}  among others. Regularization  involves formulating the required restored image ${g}_{opt}$ as a minimizer of a cost function involving the 
 observed image   ${m}$.  
 The cost is   formulated as    a sum of a data fitting term $G({g,m})$  and a regularization term $R({g})$ resulting in 
 the following minimization problem:
 \begin{equation} \label{Equation  : Regularization}
 	{{g}}_{opt}  	 = \argmin{{g}}{G({g,m} )  + \lambda  R({g})}
 \end{equation} 
 where $\lambda$ is the regularization parameter which controls the relative weighting of regularization term  against the data fitting term in the overall cost. 
 The choice of data fitting term $G(g,m)$  is dependent on the image formation forward-model.  The regularization term enforces any prior we have about the class 
 of images we are trying to restore often in the form of some image regularity condition.

 Regularization functional $R(g)$ is  primarily designed based on study of image statistics, while some are  data driven designs as observed in learning based
 methods.  
 Image statistics often depend on imaging modality and the type of objects being imaged. One of the earliest successful regularization function for restoration of natural images is Total Variation. Total Variation,  more specifically, first-order total variation (TV-1)  is defined as the sum of absolute values of image derivatives along image directions.  It works on the principle that discrete natural images have a limited  amount of variation which is captured via the first-order derivatives along the two image dimensions.  For bio-medical imaging modalities often a better choice is second-order Total Variation (TV-2)  where the sum of second-order derivatives along the two image dimensions are employed instead of the first order ones.  Replacing first-order derivatives with the second-order ones leads to more natural  intensity variation in the 
 reconstructed images  \cite{lefkimmiatis2011hessian}. A generalization of TV-2 known as Hessian-Schatten regularization (HS) was successfully employed for the restoration of biological images \cite{lefkimmiatis2013hessian}. HS regularization employed Schatten norm on the image Hessian. HS regularization and its variants  have been successfully applied in a wide variety of imaging forward models including natural images and biological images.
 
 Further developments led to more complex forms of regularization.  Among these, two  forms  are prominent in the literature:  (i) sum of norms regularization \cite{lindsten2011clustering} and  (ii) infimal decomposition-based regularization \cite{holler2014infimal}.   
 Both approaches multiple types of derivatives, but they are fundamentally different in the way the multiple order are combined, which leads to a significant difference in the performance. 
 In the first approach,  $R(g)$  takes  the form:  
 \begin{equation}
 	\label{eq:reg2t}
 	R(g)=\lambda_1 R_1(g) + \lambda_2 R_2(g)
 \end{equation}  
 where $R_1$ and $R_2$ are the functions capturing different priors based on different types of derivatives. 
 This will produce a  reconstruction that balances   regularity assumptions enforced by $R_1$ and $R_2$ [COTV].
 Unfortunately, this balancing is global in the sense that we can control only the overall agreement of the solution
 to the priors captured by the terms  $R_1$ and $R_2$.    Real images have non-stationary statistics,  and hence
 it will be more advantageous to combine different terms in  a spatially adaptive manner.  This can be accomplished
 by, for example,  making  the weights $\lambda_1$ and $\lambda_2$ spatially varying, i.e.,  by making it vary from
 pixel to pixel.  In this case,  the weights  $\lambda_1$ and $\lambda_2$ themselves become images and     determining
 these images along with  the required image becomes a challenge.    This problem can also be handled by
 an auxiliary regularization \cite{viswanath2020image}.      The main problem in this approach is that the overall cost becomes
 non-convex.

 The second approach is based on  infimal convolution 
 \cite{holler2014infimal}.   Here  the regularization functional $R(g)$ itself  is defined  by a minimization problem.
 Infimal convolution based regularization functional $R(g)$ takes the following form: 
 \begin{equation}
 	\label{eq:reginf}
 	R(g) = \mino{v_1 +v_2=g	}{R_1(v_1) +R_2(v_2)}  
 \end{equation}
 Here  $v_1$ and $v_2$ are  auxiliary variables satisfying the condition that $v_1+v_2=g$.  Note that here too,  $R(g)$ is defined as a sum of two functions $R_1$ and $R_2$. 
 But this time, $R_1$ acts on a variable $v_1$ and $R_2$ acts on a variable $v_2$ where $v_1+v_2=g$.   In the presence of data-fitting cost   that is defined on the sum
 $g=v_1+v_2$,  minimization with respect to  $v_1$   and $v_2$  leads to spatial adaptivity;    the relative influence of  $R_1$ and $R_2$      is determined adaptively  in a
 spatially varying manner depending on the spatial structure of the input image $m$.   Hence we get an effect that is similar to the method of \cite{viswanath2020image}
 with an advantage that the cost    here is convex.    An additional advantage is that  this formulation has a  higher extent of adaptivity:     it is possible to get
 a solution in which  either $R_1$ or  $R_2$ has  no influence at all  in some  spatial locations  depending on the  local structure in $m$  such that overall distortion is minimized.
 Total Generalized Variation (TGV) \cite{guo2014new,bredies2010total} is the most popular regularization designed using this concept and is used successfully in restoration of MRI images.

 More recently learning based methods have been applied successfully to restoration of images from a wide variety of imaging modalities. These belong to the class of deep learning based methods where the image prior is represented by using Convolution Neural networks. These methods use end to end learning based methods or use of CNN networks as a prior for  
 regularization-based  image restoration \cite{ulyanov2018deep}. In microscopy it has been applied in restoration of 2D images. It has also been applied to other image analysis problems in microscopy as well. \cite{xing2017deep,von2021democratising,liu2021survey}. The disadvantages of using learning in microscopy image analysis has been discussed in \cite{hoffman2021promise}.
 
 A  related category of inverse problems is  restoration of spatio-temporal 2D images observed over time. Even though a spatio-temporal image
 is simply a sequence of 2D images, the following are the disadvantages of applying 2D image restoration for each time-point independently:
 (i)  temporal correlation is ignored leading to poorer quality of restored images; (ii)  temporal continuity is lost in the restored images \cite{holler2014infimal}. 
 Due to the  inherent need for  imposing regularity  in both time and space, spatio-temporal image restoration  demands for the  employment of sum of norms or 
 infimal decomposition  based construction of the regularization functional.    In the sum-of-norms based construction,    a weighted sum of two terms as given in the 
 equation \eqref{eq:reg2t}   is
 typically  used [ref]:   (i)    the first term applies spatial regularization on each time point of the image sequence;  (ii)  the second term enforces temporal smoothing for each pixel independently.
 We will call such construction the combined spatio-temporal regularization (CST).  
 In infimal decomposition approach,  a form as given in the equation \Cref{eq:reginf}   is used.   Here  both the terms use spatial and temporal derivatives,  but, they
 differ by the relative weight applied to the temporal derivatives.
 This method was called the 
 Infimal Convolution Total Variation (ICTV), and we will refer this methods as ICTV-2DT.
 ICTV-2DT regularization has been successfully applied in biomedical imaging modalities such as dynamic MRI\cite{schloegl2017infimal} and SPECT imaging \cite{zhang2018infimal}.  
 Among other regularization based approaches, low rank sparse decomposition based model   for dynamic MRI reconstruction \cite{tremoulheac2014dynamic} is the most well known.    In MRI more recently deep learning based methods have been proposed. For restoration of real world video signals deep neural network based algorithms have been proposed. More recently RNN networks had been proposed for dynamic MRI reconstruction \cite{chen2022pyramid}.

 In this work,  we develop  a novel spatio-temporal regularization  based on infimal convolution employing two terms as done in ICTV-2DT.
 However, our regularization significantly differs from ICTV-2DT in how spatial and temporal derivatives as 
 distributed among the two terms.  We construct  the form of each term to make the overall regularization suitable for restoration of 2D fluorescence images recorded as a function of time.
 We call the novel regularization the Spatio-Temporal Adaptive Infimal Convolution (STAIC)  regularization.
 We also develop a computational algorithm for dynamic image restoration by using the STAIC regularization,
 and demonstrates the usefulness of our method using simulation experiments. The paper is organized as follows. In section \Cref{Section : Notation}, we introduce the notations and mathematical preliminaries. In section \Cref{Section :Related Work}, we describe the important existing  regularization approaches to deal with spatio-temporal signal restoration problems.
 In section \Cref{Section: STAIC Introduction},  we introduce the proposed STAIC regularization. In section \Cref{Section : Reconstruction Problem},  we present the STAIC regularized reconstruction as a convex optimization problem, and describe how we solve the optimization problem using the ADMM approach.
 In section \Cref{Section : ADMM subproblems},   we solve  the ADMM sub-problems that arise out of the reconstruction problem in more detail. Finally in \Cref{Section : Experiments}, we validate the utility of STAIC regularization.  
 
 \section{Notations and mathematical preliminaries}{\label{Section : Notation}}
 
 \begin{enumerate}
 
 	\item Images are represented by lower case English alphabets. For example $g$.
 	
 	\item
 	In the discussion we will use the idea of vector valued images  often refereed to as vector images.  Vector images are discrete 2D arrays where each pixel
 	location has a vector quantity. It is denoted by lower-case bold-faced letter
 	with a bold-faced lower-case letter as an argument.  For example,
 	${\mathbf v}({\mathbf r})$  is a vector image with ${\mathbf r}=[x\;y]^\top $ representing a 2D pixel
 	location. Depending on the context, the symbol denoting the pixel location
 	may be omitted.
 
 	\item
 	For a vector image, 	${\mathbf v}({\mathbf r})$,  $\|{\mathbf v}\|_{1,2}$
 	denotes  $\|{\mathbf v}\|_{1,2}=\sum_{\mathbf r}\|{\mathbf v}({\mathbf r})\|_{2}$. It is
 	the  sum of  pixel-wise $l_2$ norms, where $\sum_{\mathbf r}$ denotes the sum
 	across pixel indices. The bound of sum is the first to last pixel location in this notation.
 	The norm $\|{\mathbf v}\|_{1,2}$ is a composition of  norms and is often refereed to as a mixed norm.
  
 	\item In a scalar image having multiple frames, we use the subscript notation to refer to a particular frame number. Example $g_i$ refers to frame number $i$ of the spatio-temporal image $g$.
 	\item Index $\mathbf{r}$ is used to refer to a spatio-temporal image (2DT). Index $\mathbf{k}$ is used to refer to a 2D image. Let $\delta(\*k)$
 	and $\delta(\*r)$ represent 2D and 3D Kronecker delta respectively.
 	\item $\|\mathbf{g}\|_{1,\kappa} =  \sum_{\mathbf{r}} \|\mathbf{g}(\mathbf{r}) \|_{\kappa}$ 
 	with the definition of 
 	$\small \|\mathbf{y}\|_{\kappa} =   \sqrt{\kappa^2 (y_1^2 +y_2^2) +y_3^2}$.  
 	Further we have that 
 	$\|\mathbf{g}\|_{1,1/\kappa} =  \sum_{\mathbf{r}} \|\mathbf{g}(\mathbf{r}) \|_{1/\kappa}$ 
 	and
 	$\|\mathbf{x}\|_{1/\kappa} =   \sqrt{ y_1^2 +y_2^2+ \kappa^2 y_3^2}$.  Here $\kappa$ is a parameter.
 	\item   $\bar{d}_{xx}(\*k),\bar{d}_{xy}(\*k),\bar{d}_{yy}(\*k)$ are discrete filters implementing 2D second order derivatives. In addition  $d_{xx}(\*r),d_{xy}(\*r), d_{xt}(\*r)$,$d_{yy}(\*r),d_{yt}(\*r),d_{tt}(\*r)$ are   discrete filters implementing the 3D second order derivatives.   \item   $M(\*r)$ is the 3D Hessian operator  encapuslating   all the second order derivative operators where the directions are $(x,y,t)$
 \end{enumerate}

 \section{Related methods in detail}\label{Section :Related Work}
 Before discussing related work, it is important to understand the expected behavior from an ideal spatio-temporal regularization function. In regions with motion, regularization function must be able to discourage temporal smoothing. In regions without motion, temporal smoothing must be promoted. All existing approaches to regularization for this inverse problem is trying to achieve this ideal behavior in some way or the other. This is possible only if the regularization function $R(g)$ is powerful enough to  
 differentiate between regions with motion and regions without motion. This points towards the need to incorporate both spatial and temporal derivatives in the design of regularization function. The presence of temporal derivatives also ensures that such functions operate  differently to  classical image regularization methods when employed in the resultant optimization problem.

 Early methods that consider temporal derivatives together with spatial derivatives constructed the overall regularization as a simple weighted sum of  functionals constructed using spatial and temporal derivative independently.
 This sum of norm based spatio-temporal regularization  which we called Combined Spatio-Temporal (CST) is defined as :
 \begin{equation}
 	R_{CST}(g)=\|\mathcal{M}*{g}\|_{1,2} +\|d_{tt}*{g}\|_{1,2}
 \end{equation}
 where $\mathcal{M}$ is the second order Hessian operator in 3D and $D_{tt}$ is second order derivative along time dimension. In restoration of spatio-temporal signals, $R_{CST}$ is known not to satisfy the desirable properties of an ideal spatio-temporal regularization stated above. 
 One disadvantage of  the above formulation is that it is blind to  the fact that characteristics of motion varies spatially and temporally,   and there is no spatially and temporal varying trade-off between spatial and temporal derivatives. 
 In regions with motion both components will have nontrivial contribution to overall cost leading temporal smoothing  which is undesirable.  In regions without motion spatial smoothing occurs in addition to temporal smoothing which is again undesirable.This shortcoming prompted design of spatio-temporal regularization employing the infimal decomposition approach as done in the ICTV-2DT method introduced before.
 Let $g(\*r)$  be the discrete candidate spatio-temporal (2D+time) image, where $\*r=[x\;y\;t]^\top$ is the discrete 3D pixel index. Assume that  image $g(\mathbf{r})$ is of dimensions $n_1 \times n_2 \times n_F$ where $n_F$ is the number of frames (2D images). Hence each frame is of dimension $n_1 \times n_2$. The regularization is constructed in the form of minimization as given below:
 \begin{equation} 
 	R_{ICTV}(g) = 	\operatornamewithlimits{min} \limits_{v}\|{\cal S}_{\kappa_1}\nabla(g-v)  \|_{1,\kappa_1}  + 
 	\|{\cal S}_{\kappa_2}\nabla(v) \|_{1,\kappa_2},
 \end{equation} 
 where ${\cal S}_{\kappa}=diag(\kappa,\kappa,1)$ and 
 $\nabla$ is the  3D gradient operator. 
 Here   $\kappa_1$ and $\kappa_2$  are  user  parameters that  determines relative weight between spatial and temporal derivatives. 
 It may be noted that  the method requires tuning for 
 $\kappa_1$  and $\kappa_2$   in addition to the tuning of the overall weight $\lambda$. 
 This form has the advantage over CST  in the same way as the form introduced in the equation \eqref{eq:reg2t}   has an advantage over the form in the equation  
 \eqref{eq:reginf}. In other words,  combining two different types of sub-functional in the form of infimal convolution is always advantageous over a simple weighted.
 However,  ICTV-2DT  still has a disadvantage. 
 This can be  understood by analyzing its effect on regions with motion and without motion. 
 In regions with motion,   it will be desirable to eliminate temporal smoothing completely, but, ICTV-2DT  does not do so.
 Similarly, in regions without motion,  it will be desirable to eliminate spatial smoothing completely,  but,  ICTV-2DT  does not do so.
 The reason is that both sub-functional have both temporal and spatial derivatives and, they differ only by  the relative weight between spatial
 and temporal derivatives. Further, it uses first order derivatives which is again undesirable.

 \section {Spatio-Temporal Adaptive Infimal Convolution (STAIC) Regularization for  2DT image restoration}
 \label{Section: STAIC Introduction}
 
 We propose STAIC regularization   for spatio-temporal  images by   removing the disadvantages of ICTV-2DT while strengthening the design to cater
 to unique features of  2DT  fluorescence  images.    Let $g_i(\*k)$  represent the $i^{th}$ frame of the spatio-temporal image $g(\*r)$   where $\*k=[x\;y]^\top$  is the 
 discrete $2D$ pixel index. Consider two  linear operators $A(\*k)$ and $M(\*r)$ defined as follows :
 \begin{equation*}
 	A(\*k)=\begin{bmatrix}
 		\bar{d}_{xx}(\*k)   \\    
 		\bar{d}_{xy}(\*k)   \\
 		\bar{d}_{yx}(\*k)      \\
 		\bar{d}_{yy}(\*k)   \\
 		\delta(\*k)
 	\end{bmatrix}, 	M(\*r)=\begin{bmatrix}
 		d_{xx}(\*r)       & d_{xy}(\*r)  & d_{xt}(\*r) \\
 		d_{yx}(\*r)       & d_{yy}(\*r)  & d_{yt}(\*r) \\
 		d_{tx}(\*r)       & d_{ty}(\*r)  & d_{tt}(\*r)  
 	\end{bmatrix}
 \end{equation*}
 Next, we define the following components, which act on 2DT images:
 \begin{align}
 	\label{eq:R1def}
 	\bar{R}_1(u_i) & = \|A*u_i   \|_{1,2} \\
 	\label{eq:R2def}
 	R_2(v) & = \|(M*v) \|_{1,F}
 \end{align}
 Here $u$  and $v$ are 2DT images, and $u_i$  is the $i^{th}$ 2D time frame of    of the 2DT image $u$.     Note that
 $ \|A*u_i   \|_{1,2} =  \sum_{\bf k}   \|(A*u_i)({\bf k}) \|_2 
 =  \sum_{\bf k}   \sqrt{  ((\bar{d}_{xx}*u_i)(\*k))^2 +  2((\bar{d}_{xy}*u_i)(\*k))^2  +  ((\bar{d}_{yy}*u_i)(\*k))^2 + (u_i({\bf k}))^2 }$. 
 This is a modified 2D second order TV applied on $u_i$, where the modification is simply the addition of the intensity term. 
 Next,  we note that  $R_2(v)$  computes  the 3D  second-order TV   applied on the 2DT sequence $v$.

 The proposed  STAIC regularization is defined as follows :
 \begin{equation*}
 	S(g,\alpha_s,\alpha_t) = \mino{v}{ \alpha_s R_1(g,v) +\alpha_t R_2(v)  } 
 \end{equation*}
 where $R_1(g,v) = \sum_{i=1}^{n_F}\bar{R}_1(g_i-v_i)$. The user defined parameters $\alpha_s$, $\alpha_t$ helps control the relative strength of $R_1( \cdot, \cdot )$ and $R_2( \cdot)$ in the regularization function $S(g,\alpha_s,\alpha_t)$.

 Despite  the role of $v$ as an auxiliary variable in the definition of regularization, the component $v$ is regularized to have finite amount of variations in its 3D structure. This is achieved by  constraining the Frobenius  norm  of 3D second order derivatives. This reflects our desire to ensure that $v$ also resemble a realistic spatio-temporal signal.
 In the first term $R_1(g,v)$, we put a constraint on spatial regularity of the difference signal by incorporating   an intensity term in addition to the spatial derivative term. This combination of derivatives and pixel intensity  promotes the presence of blob-like structure in the restored image. Such image features are often encountered in fluorescence microscopy images which is desirable in our setting due to the TIRF microscopy origin of our signal of interest.

 To understand the effect of the two terms on the overall regularization function, it is important to consider  regions with motion and without motion separately.  
 \begin{itemize}
 	\item In regions with motion, $v$  will tend to be zero, and so  only  $R_1$ comes into play;  this means that 
 	there will be negligible  temporal smoothing, which will reduce motion blurring
 	\item In regions without motion,   $v$ will  tend to be close to  $g$ so that  only  	$R_2$ come into play;  
 	since $R_2$ imposes 3D smoothing, this lead to  robust noise removal                
 \end{itemize}
 It may be noted that the design of $R_1$ in STAIC removes an important drawback of ICTV-2DT in regions with motion. STAIC due to the presence of only spatial derivatives in its $R_1$ term discourages temporal smoothing completely in regions with motion. 
 This is an important feature as it avoids motion blurring  in the final image estimate. In addition, a crucial design change is the presence of a pure intensity term $g-v$ in the $R_1$ term. This is done due the following reasons :
 
 \begin{itemize}
 	\item   We target our method of restoration fluorescence images;   fluorescence microscopy   images  often have images exhibiting a blob like pattern which is promoted by the presence of intensity term  
 	\item In addition the intensity term help promote sparsity of regions with motion in turn helping remove micro motions which are often the artifacts caused by Poisson nature of noise. This helps in recovering all large and small motions while removing any spurious motion in the signal.
 \end{itemize} 
 In addition, STAIC formulations also do not employ any parameter to control the relative strength of spatial and temporal derivatives by virtue of its design leaving us one less parameter to tune. In summary,(i) the redesigned  design of purely spatial term $R1$ (ii) intensity augmentation of $R1$ helps us achieve the goal of designing a regularization scheme that  has close to ideal spatio-temporal behavior in regions with motion and without it with limited number of parameters. This ensures that the   reconstruction  is  closer to the ground truth in terms of both spatial and temporal features.

 \section{Signal Reconstruction Using STAIC regularization}  
 \label{Section : Reconstruction Problem}

 \subsection{The Cost Function}
 Let $h$ denote the 2D impulse response of the  microscope. The noise model in fluorescence images  is modeled to be  mixed  Poisson-Gaussian where a signal corrupted by Poisson noise is  subject to an additive Gaussian noise. The generation of measured image $m$ is therefore  modeled as follows:
 \begin{equation*}
 	m = \mathcal{P}(h * g)  + \eta 
 \end{equation*} 
 where $\mathcal{P}$ is the  operator representing Poisson process, $\eta$ represents the additive Gaussian noise, $m$ represents the observed noisy blurred image and $g$ represents the ground truth image. The data fitting term is often chosen to be the maximum likelihood estimate of the noise model. But    we choose the least squares data fitting term due to ease of optimization \cite{arigovindan2013high}. The data fitting term is 
 \begin{equation}
 	\label{Equation : datafitting }
 	G(h,g,{m})=\frac{1}{2}\sum_{i=1}^{n_F}\|(h*g_i)-m_i\|_{F}^2
 \end{equation}
 It may be noted that the blurring happens frame wise as we are considering a 2D signal observed over time. The STAIC regularized image reconstruction optimization problem can be expressed as
 \begin{equation}
 	g_{opt} =  \argmin{g}{G(h,g,m) +S(g,\alpha_s,\alpha_t)}
 \end{equation}
 By observing that the regularization functional  $S(g,\alpha_s,\alpha_t) $  is defined as an optimization problem over variable $v$, we can express the reconstruction problem as:
 \begin{eqnarray}  
 	\label{Equation;2variables} 
 	&	(g_{opt},v_{opt}) =  \argmin{g,v}{G(h,g,m) +\bar{S}(g,v,\alpha_s,\alpha_t)}\\  &\text{ where } 
 	\bar{S}(g,v,\alpha_s,\alpha_t) =  \alpha_s R_1(g,v) +\alpha_t R_2(v) \nonumber
 \end{eqnarray} 
 resulting in an optimization problem in 2 variables $g$ and $v$.  This is possible because, the variable $v$ do not appear in the data fitting term enabling it to be moved to the outer optimization step. 
 We introduce a variable (vector image)  $\mathbf{f(\*r)}=[g(\*r) \;\;v(\*r)]^\top$ concatenating the two variables in the above optimization problem to enable a simpler restatement of the cost of optimization. To this end we define the following new linear operators :
 \begin{align*}
 	{\mathbf h} & = [h(\mathbf{k})\;\;0 ],  \\
 	{\mathbf T}_s & = \left[\begin{array}{cc}
 		d_{xx}(\mathbf{k}) & 0   \\ 
 		d_{xy}(\mathbf{k}) & 0  \\
 		d_{yx}(\mathbf{k}) & 0  \\
 		d_{yy}(\mathbf{k}) & 0  \\
 		\delta(\mathbf{k})    & 0 \\
 		0         &  d_{xx}(\mathbf{k}) \\
 		0         &  d_{xy}(\mathbf{k}) \\	
 		0         &  d_{yx}(\mathbf{k}) \\	
 		0         &  d_{yy}(\mathbf{k}) \\
 		0         &   \delta(\mathbf{k})
 	\end{array}   \right]\text{ and }  \;\; 
 	{\mathbf T}_t=\left[ \begin{array}{cc}
 		0 & d_{xx}(\*r)   \\ 
 		0 & d_{yy}(\*r)   \\
 		0 & d_{xy}(\*r)   \\ 
 		0 & d_{yx}(\*r)   \\ 
 		0 & d_{xt}(\*r)   \\
 		0 & d_{tx}(\*r)   \\
 		0 & d_{yt}(\*r)   \\
 		0 & d_{ty}(\*r)   \\
 		0 & d_{tt}(\*r)  
 	\end{array} \right] 
 \end{align*}
 
 \begin{align} \label{Definition : Matrix A_s}
 	&{A}_s =  \small \left[ \begin{array}{cccccccccc}
 		1/\sqrt{2} & 0 & 0 & 0 &0 & -1/\sqrt{2} & 0 & 0 & 0  & 0 \\
 		0&1/\sqrt{2} & 0 & 0 & 0 &0 & -1/\sqrt{2} & 0 & 0& 0    \\
 		0&0&1/\sqrt{2} & 0 & 0 & 0 &0 & -1/\sqrt{2} & 0& 0   \\ 
 		0&0&0&1/\sqrt{2} & 0 & 0 & 0 &0 & -1/\sqrt{2} & 0 \\
 		0&0&0&0&1/\sqrt{2} & 0 & 0 & 0 &0 & -1/\sqrt{2} 
 	\end{array}  \right]
 \end{align}
 The optimization problem in \eqref{Equation;2variables} can now be reformulated in terms of $\mathbf{f}$ only as :
 \begin{align}\small \label{Equation : Unconstrained STAIC Cost}
 	J(\*f,\alpha_s,\alpha_t) = &\frac{1}{2}\sum_{i=1}^{n_F}\|\*h*\mathbf{f}_i-m_i\|_{F}^2+ 
 	\sqrt{2} \alpha_s\sum_{i=1}^{n_F}\sum_{\mathbf{k}}\|A_s\big((\mathbf{T_s}*\*f_i)(\mathbf{k})\big)\|_{2} +\alpha_t  \sum_{\*r}\|(\mathbf{T_t}*\*f)(\*r)\|_{2} +	\mathcal{B_C}(\*f) 
 \end{align}
 Here, 
 the  additional term $	\mathcal{B_C}(\cdot)$ is  added to impose bound constraint on $\*f$ which restrict the range of pixel values $\*f$ can take. This helps impose commonly used non-negativity condition on pixel values. Here $\mathcal{B_C}(\cdot)$ is defined as follows 
 \begin{equation}
 	\mathcal{B_C}(\mathbf{f}) =
 	\begin{cases}
 		\text{0} &\quad\text{if } \*f \in \mathcal{C} \\
 		\infty &\quad\text{otherwise} 
 	\end{cases}
 \end{equation}

 Here $\mathcal{C}=\{\*f \;|\; lb \leq \*f(\r) \leq ub \;\; \forall \*r\}$ where  $lb$ and $ub$ represents the smallest and highest permitted pixel values. The new image restoration optimization problem over variable $\mathbf{f}$ now takes the form :
 \begin{equation} \label{Equation : Original Optimization  Problem}
 	\mathbf{f}_{opt} =  \argmin{\*f}{ 	J(\*f,\alpha_s,\alpha_t)}
 \end{equation}
 \subsection{Proposed ADMM Method}
 The cost   $J(\*f,\alpha_s,\alpha_t)$  to be minimized is a convex  function in variable $\mathbf{f} $. We propose to design an algorithm to minimize the cost by employing the Alternating Direction Method of Multipliers (ADMM) \cite{boyd2011distributed} algorithmic framework. ADMM framework is most suited for minimizing convex cost under linear equality constraints. Our original optimization problem in \eqref{Equation : Original Optimization  Problem} can be  reformulated to get an equivalent  linearly constrained convex problem. This is done by  by introduction of new variables $w_m , \mathbf{w}_s,\mathbf{w}_t$ and $\mathbf{w}_b$ and as a result, the updated problem becomes:
 \begin{align*}
 	&\operatornamewithlimits{min}\limits_{\mathbf{f},w_m,\*w_s,\*w_t,\*w_b}{\;\;\frac{1}{2}\sum_{i=1}^{n_F}\|w_{m_i}-m_i\|_{F}^2}  +  \sqrt{2} \alpha_s\sum_{i=1}^{n_F}\sum_{\mathbf{k}}\|A_s \big(\mathbf{w}_{s_i} (\mathbf{k})\big) \|_{2}+\alpha_t \sum_{\*r}  \|\mathbf{w}_t(\*r)\|_{2} +	\mathcal{B_C}(\mathbf{w}_b)\\
 	&\text{ subject to } \quad \quad
 	w_{m_i}=\mathbf{h}*\mathbf{f}_i,\;\;\mathbf{w}_{s_i}=\mathbf{T}_s*\mathbf{f}_i,\;\;\mathbf{w}_t=\mathbf{T}_t*\mathbf{f},\;\;\mathbf{w}_b =  \*f
 \end{align*}
 
 It may be noted that we have converted the unconstrained problem in \eqref{Equation : Original Optimization  Problem} to a constrained optimization form where all constraints are  linear equality constraints.  To allow a simpler algorithm statement, we introduce a combined operator $\mathbf{T}$ and a combined vector $\*w$ defined as follows :
 
 \begin{align}
 	&\*T  = \begin{bmatrix}
 		\*h(\mathbf{k}) \\
 		\*T_s(\mathbf{k}) \\
 		\*T_t(\*r) \\
 		\*e(\*r)
 	\end{bmatrix}, \quad
 	\*w=\begin{bmatrix}
 		w_m \\
 		\*w_s \\
 		\*w_t \\
 		\*w_b
 	\end{bmatrix}
 \end{align}
 where $\mathbf{e}(\*r)=[\delta(\*r) \;\; \delta(\*r)]$. Under this definition, $\mathbf{e} * \*f = \*f $.
 The linearly constrained problem can now  be stated in a compact form as : 
 \begin{align}
 	&	(\mathbf{f}^*,\mathbf{w}^*) = \argmin{\*f,\*w}{R(\*w,\alpha_s,\alpha_t)}  \\
 	& \text{ subject to } \quad \quad\mathbf{T}*\*f = \*w   \nonumber
 \end{align}   
 where $R(\*w,\alpha_s,\alpha_t)=G(w_m,m)+ \sqrt{2}\alpha_s\sum_{i=1}^{n_F}\sum_{\mathbf{k}}\| {A}_s(\mathbf{w}_{s_i}(\mathbf{k}))\|_{2} +\alpha_t \sum_{\*r} \|\mathbf{w}_t(\*r)\|_{2} +	\mathcal{B_C}(\mathbf{w}_b)$
 and $G(w_m,m)=\frac{1}{2}\sum_{i=1}^{n_F}\|w_{m_i}-m_i\|_{F}^2$.
 The next step in ADMM framework is to construct the Augmented Lagrangian \cite{boyd2011distributed} $\mathcal{L}(\*f,\*w,{\pmb \beta},\alpha_s,\alpha_t)$   of the above linearly constrained cost as:
 \begin{equation} \small
 	\mathcal{L}(\*f,\*w,\pmb{\beta},\alpha_s,\alpha_t) = R(\*w,\alpha_s,\alpha_t) + \langle {\pmb \beta}, \mathbf{T}*\mathbf{f}-\mathbf{w}\rangle +\frac{\rho}{2}\|\mathbf{T}*\mathbf{f}-\mathbf{w}\|_2^2
 \end{equation}
 where ${\pmb \beta} $ is the Lagrangian multiplier and $\rho$ is a user supplied ADMM parameter. Here the dimensions of $\pmb \beta$ is same as that of $\*w$. Finally, ADMM algorithm involves collection of alternative minimization of sub-problem with respect to variables $\*f$ and $\*w$ followed by an update step involving variable $\pmb \beta$. Assume that $\*f^{(k)},\*w^{(k)} ,{\pmb \beta}^{(k)}$ are the current estimates , the ADMM algorithm involves the following steps
 \begin{align}   		  		
 	\*w^{(k+1)}&=\operatornamewithlimits{argmin}\limits_{\*f}{{L}(\*f^{(k)},\*w,{\pmb \beta}^{(k)},\alpha_s,\alpha_t)} \label{Equation : w-subproblem}\\
 	\*f^{(k+1)}&=\operatornamewithlimits{argmin}\limits_{\*f}{L(\*f,\*w^{(k+1)},{\pmb \beta}^{(k)},\alpha_s,\alpha_t)} \label{Equation : f-subproblem}\\
 	\text{ and } \;\;{\pmb \beta}^{(k+1)}& = {\pmb \beta}^{(k)} + \rho\big(\mathbf{T}*\mathbf{f}^{(k+1)}-\mathbf{w}^{(k+1)}\big) \label{Equation : lambda-subproblem}
 \end{align}
 The first two equation involves solving two optimization problems over variables $\mathbf{f}$ and $\mathbf{w}$ respectively which is discussed next.
 
 \section{Solving the Sub-problems of ADMM Algorithm} \label{Section : ADMM subproblems}
 
 We will now discuss how the  sub-problems in  \eqref{Equation : w-subproblem}, and \eqref{Equation : f-subproblem} are solved to obtain the intermediate variables $\*w^{(k+1)} $ and 
 $\*f^{(k+1)}$ that appear in the ADMM iterative scheme.
 \subsection{The w problem}
 The sub-problem in \eqref{Equation : w-subproblem}  can be equivalently simplified to the following form    
 \begin{align}
 	&	L_{w,k}(\*w,\alpha_s,\alpha_t)=  R(\*w,\alpha_s,\alpha_t) + \frac{\rho}{2}\|\*w-\bar{\*x}^{(k)}\|_2^2 \\
 	&\quad \text{ where }  \quad 
 	\bar{\mathbf{x}}^{(k)} = \mathbf{T}*{\*f^{(k)}} + \frac{1}{\rho}{\pmb \beta^{(k)}}
 \end{align}
 For cleaner presentation of sub-problems, we introduce the notation $\*x=\bar{\*x}^{(k)}$ and $\hat{\*w}=\*w^{(k+1)}$.  Since $\*w$ is made up of sub vectors $w_m,\*w_b,\*w_t,\*w_s$, we separate the above problem into sub-problems involving constituent variables.
 
 \begin{align}
 	\label{eq:wmprob2}
 	\mbox{$w_m$-prob.:} \; &  \hat{w}_m = 
 	\underset{w_m}{\operatorname{argmin}} 
 	\underbrace{\frac{\rho}{2} \|x_m-w_m\|_{2,2}^2 + 
 		G({w}_m,   {m})}_{\bar{L}_{m}(w_m,x_m)}  \\
 	\label{eq:wbprob2}
 	\mbox{$\*w_b$-prob.:} \;\;\;&  \hat{\*w}_b = 
 	\underset{\*w_b}{\operatorname{argmin}} \;\; 
 	\underbrace{\frac{\rho}{2} \|\*x_b-\*w_b\|_{2,2}^2 + 
 		{\mathcal{B_C}}(\*w_b)}_{\bar{L}_{b}(\*w_b,\*x_b)}  \\
 	\label{eq:wfprob2}
 	\mbox{${\mathbf w}_t$-prob.:} \;\;\; &   \hat{\mathbf w}_t = 
 	\underset{{\mathbf w}_t}{\operatorname{argmin}} \;\; 
 	\underbrace{\frac{\rho}{2} \|{\mathbf x}_t-{\mathbf w}_t\|_{2,2}^2 + 
 		\alpha_t \| \mathbf{w}_t\|_{1,2}}_{\bar{L}_{t}({\mathbf w}_t, {\mathbf x}_t, \alpha_t)} \\
 	\label{eq:wsprob2}
 	\mbox{${ \*w}_s$-prob.:} \;\;\; &  \hat{  \*w}_s =  
 	\underset{{\mathbf w}_s}{\operatorname{argmin}} \;\; \sum_{i=1}^{n_F}
 	\underbrace{\frac{\rho}{2} \|{\mathbf x}_{s_i}-{\mathbf w}_{s_i}\|_{2,2}^2 + 
 		\sqrt{2}\alpha_s  \sum_{\mathbf{k}}\|{A}_s (\mathbf{w}_{s_i}(\mathbf{k}))\|_{2}}_{\bar{L}_{s_i}({\mathbf w}_{s_i}, 
 		{\mathbf x}_{s_i}, \alpha_s)} 
 \end{align}
 \subsubsection{Decomposing problems pixel-wise}
 The cost $\bar{L}_{m}(w_m,x_m)$ is separable across pixels as shown below:
 
 \begin{align}
 	\bar{L}_{m}(w_m,x_m)  = & 
 	\frac{\rho}{2} \|x_m-w_m\|_{2,2}^2 + 
 	G({w}_m,m)  \\
 	= &   \frac{\rho}{2} \|x_m-w_m\|_{2,2}^2 +    \frac{1}{2}\| w_m -m\|_{2}^2\\
 	=&\sum_{\*r} \underbrace{ \frac{\rho}{2} (x_m(\*r)-w_m(\*r))^2 +    \frac{1}{2}(w_m(\*r) -m(\*r))^2}_{{L}_{m}(w_m(\r),x_m(\r))}
 \end{align}

 Hence  the pixel wise cost ${L}_{m}(w_m(\r),x_m(\r))$ is given by:
 \begin{equation*}
 	{L}_{m}(w_m(\r),x_m(\r))  =     \frac{\rho}{2} (x_m(\r)-w_m(\r))^2   
 	+   \frac{1}{2}(w_m(\r)-m(\r))^2
 \end{equation*}

 The cost function $\bar{L}_{b}(w_b,x_b)$ is separable across 3D pixel index $\*r$ because
 $\mathcal{B_C}(\mathbf{w}_b)= \sum_{\*r} \bar{ \mathcal{B_C}}(\mathbf{w}_b(\*r)) $ where  
 \begin{equation}
 	\bar{ \mathcal{B_C}}(\mathbf{w}_b(\*r)) =
 	\begin{cases}
 		\text{0} &\quad\text{if } \mathbf{w}_b(\*r) \geq 0\\
 		\infty &\quad\text{otherwise} 
 	\end{cases}
 \end{equation}
 The cost function reformulated as a sum over pixels can be stated as:
 \begin{equation}
 	\bar{L}_{b}(w_b,x_b)   = \sum_\*r 
 	\underbrace{\frac{\rho}{2} (\mathbf{x}_b(\*r)-\mathbf{w}_b(\*r))^2 + 
 		\bar{ \mathcal{B_C}}(\mathbf{w}_b(\*r))}_{
 		{L}_{b}(\mathbf{w}_b(\*r),\mathbf{x}_b(\*r)) }
 \end{equation}

 Now $\bar{L}_{s_i}({\mathbf w}_{s_i}, {\mathbf x}_{s_i}, \alpha_s)$ can be expanded across pixels as follows:
 \begin{equation}
 	\bar{L}_{s_i}({\mathbf w}_{s_i}, {\mathbf x}_{s_i}, \alpha_s)  = \sum_{\mathbf{k}}
 	\underbrace{\frac{\rho}{2}\|{\mathbf x}_{s_i}(\mathbf{k})-{\mathbf w}_{s_i}(\mathbf{k})\|_{2}^2 + 
 		\sqrt{2}	\alpha_s \|A_s{  \*w}_{s_i}(\mathbf{k})\|_{{2}}}_{
 		{L}_{s}({\mathbf w}_{s_i}(\mathbf{k}), {\mathbf x}_{s_i}(\mathbf{k}), \alpha_s)},
 \end{equation}
 
 Finally, $\bar{L}_{t}({\mathbf w}_t, {\mathbf x}_t, \alpha_t)$ can also be expanded across pixels courtesy the use of mixed vector matrix norm
 
 \begin{equation} \label{Equation: ws_subproblem_pixel}
 	\bar{L}_{t}({\mathbf w}_t, {\mathbf x}_t, \alpha_t)    = \sum_\*r
 	\underbrace{
 		\frac{\rho}{2} \|{\mathbf x}_t(\*r)-{\mathbf w}_t(\*r)\|_{2}^2 + 
 		\alpha_t \| {\mathbf w}_t(\*r)\|_{2}}_{
 		{L}_{t}({\mathbf w}_t(\*r), {\mathbf x}_t(\*r), \alpha_t)  }
 \end{equation}

 We have shown so far that all  the sub-problems are separable across pixels.
 Hence the solution to the
 minimization problems of equations \eqref{eq:wmprob2},   \eqref{eq:wbprob2}, 
 \eqref{eq:wsprob2},
 and \eqref{eq:wfprob2}, can be expressed as following:
 \begin{align}
 	\label{eq:wmrobpix}
 	\hat{w}_m(\r) = & 
 	\underset{z\in \mathbb{R}}{\operatorname{argmin}} \;\;
 	{L}_{m}(z,x_m(\r)), \\
 	\label{eq:wbrobpix}
 	\hat{\mathbf{w}}_b(\r) = & 
 	\underset{z\in \mathbb{R}}{\operatorname{argmin}} \;\;
 	{L}_{b}(z,\mathbf{x}_b(\r)), \\
 	\label{eq:wsprobpix}
 	\hat{\mathbf w}_{s_i}(\mathbf{k}) = &
 	\underset{{\mathbf z}\in \mathbb{R}^{10}}{\operatorname{argmin}} \;\;
 	{L}_{s}({\mathbf z},{\mathbf x}_{s_i}(\mathbf{k}), \alpha_s)\text{ , and}  \\
 	\label{eq:wfprobpix}
 	\hat{\mathbf w}_t(\r) = &
 	\underset{{\mathbf z}\in \mathbb{R}^9}{\operatorname{argmin}} \;\;
 	{L}_{t}({\mathbf z},{\mathbf x}_t(\r), \alpha_t).
 \end{align}
 \subsubsection{{Solution to  the pixel-wise sub-problems}}
 \label{sec:subp}
 
 The solution to the $ {w}_m$ sub-problem is  obtained  by exploiting the fact that the cost  ${L}_{m}(z,x_m(\r))$ is a differentiable function. The minima is obtained by  finding the stationary point of the cost function  and the resultant optimal point $\hat{w}_m(\r)$ is:
 \begin{equation}
 	\hat{w}_m(\r)  = \frac{\rho x_m(\r)  + m(\r)}{\rho + 1}
 \end{equation}
 
 The solution to the
 $w_b$-problem  is also  simple, and it is the clipping of the pixels by bound
 that defines the set $\mathcal{C}$  \cite{parikh2014proximal}.
 The optimal point  $\hat{w}_b(\r)$  as given below:
 \begin{equation}
 	\hat{\mathbf{w}}_b(\r) =  \mathbb{P}_{ \mathcal{C}}(\mathbf{x}_b(\r)),
 \end{equation}
 where ${\mathbb{ P}}_{\mathcal{C}}(\cdot)$  denotes the operation of clipping the pixel values within the   bounds in definition of $\mathcal{C}$.

 The $\mathbf{w}_t$ sub-problem could be understood as evaluating the well known proximal operator \cite{parikh2014proximal} of $\ell_2$ norm $\frac{\alpha_t}{\rho}\|\cdot \|_2$ at the point ${\mathbf x}_t(\*r)$.  Hence the solution to the  $\mathbf{w}_t$ sub-problem is given by 
 
 \begin{align}
 	\hat{\mathbf w}_t(\r)&= 	\underset{{\mathbf z}\in \mathbb{R}^9}{\operatorname{argmin}} \;\;	\frac{\rho}{2} \|{\mathbf x}_t(\*r)-\*z\|_{2}^2 + 
 	\alpha_t \| {\mathbf z} \|_{2}\\
 	&=\begin{cases}
 		\big(1-\frac{\alpha_t}{\rho \|{\mathbf x}_t(\*r)\|_2}\big){\mathbf x}_t(\*r) & \|{\mathbf x}_t(\*r)\|_2\geq \frac{\alpha_t}{\rho}\\
 		\mathbf{0} &otherwise
 	\end{cases}
 \end{align}
 
 The $\mathbf{w}_s$ sub-problem is more complicated as its involves composition of  a linear operator with a norm function. The solution of this sub-problem is given by the following lemma:

 \begin{lemma} \label{Lemma : As proximal} Let $\mathbf{y} \in \mathbb{R}^{10}$. Let $\mathbf{y}_1\in \mathbb{R}^{5}$ be its sub-vector with first five entries and $\mathbf{y}_2\in \mathbb{R}^{5}$ be its sub-vector with last five entries. The solution of optimization problem  $\;\;\argmin{\mathbf{z} \in \mathbb{R}^{10}}{\frac{\rho}{2}\| {{  \mathbf{y}}}-\mathbf{z}\|_{2}^2 + 
 		\sqrt{2}	\alpha_s \|A_s \mathbf{z}\|_{{2}}}\;\;$  where   $A_s$ is defined in  \eqref{Definition : Matrix A_s} is given by
 	\begin{equation} 	\mathbf{z}^* = P\left[\begin{array}{c} \gamma {{\mathbf{y}}}_{1} \\ {{\mathbf{y}}}_{2}\end{array} \right] \end{equation}
 	Here    $P$  defined in \eqref{Defintion : Matrix  P of eigenvectors} is the eigenvector matrix of $A_s$  and  $\gamma =  \max(0,1 -\frac{\sqrt{2} \alpha_s}{\rho \|{{\mathbf{y}}}_{1}\|_2})$.
 \end{lemma}
 
 The proof of \Cref{Lemma : As proximal} is given in \Cref{Appendix : Proofs}. By applying   this lemma, we can conclude that the  solution of $\mathbf{w}_s$ sub-problem is as follows:
 \begin{equation}
 	\hat{\mathbf{w}}_{s_i}(\*k)= P\left[\begin{array}{c} \gamma \mathbf{x}_{s_{i1}}(\*k)  \\ \mathbf{x}_{s_{i2}}(\*k) \end{array} \right]
 \end{equation}
 
 where $\mathbf{x}_{s_{i1}}(\*k) \in \mathbb{R}^5$ is the sub-vector of $\mathbf{x}_{s_{i}}(\*k) $  with first five entries and  $\mathbf{x}_{s_{i2}}(\*k) \in \mathbb{R}^5$ is the sub-vector  of $\mathbf{x}_{s_{i}}(\*k) $ with last five entries.

 \subsection{The f sub-problem }
 \begin{equation}	
 	\*f^{(k+1)}=\operatornamewithlimits{argmin}\limits_{\*f}{L(\*f,\*w^{(k+1)},{\pmb \beta}^{(k)},\alpha_s,\alpha_t)}  
 \end{equation}

 The sub-problem in variable 	  given in  \eqref{Equation : f-subproblem}   has a simpler form once you ignore all the terms not depending on $\mathbf{f}$ in the optimization problem. The simpler from of $\mathbf{f}$ sub-problem may be stated as follows:
 
 \begin{eqnarray}
 	\*f^{(k+1)} =&\operatornamewithlimits{argmin}\limits_{\*f}\frac{1}{2}\|\mathbf{T}*\mathbf{f}-\mathbf{y}^{(k)}\|_2^2 \\\text{ where } 
 	\nonumber	& \mathbf{y}^{(k)}=\mathbf{w}^{(k+1)}-\frac{1}{\rho} \pmb{\beta}^{(k)}
 \end{eqnarray}

 For notational convenience, we let $\mathbf{y} = \mathbf{y}^{(k)}$ and $\hat{\mathbf{f}}=\mathbf{f}^{(k+1)}$.
 Recall that $\mathbf{f(\*r)}=[g(\*r) \;\;v(\*r)]^\top$From the definition of $\mathbf{T}(\*r)$ it can be observed that the cost is separable across the components $g$ and $v$ of $\mathbf{f}(\*r)$. Assume that $\mathbf{y}(\*r) =
 [ {y}_m(\*r)\;  {y}_{s,1}(\*r) \;  {y}_{s,2}(\*r)\;   {y}_{s,3}(\*r)\;   {y}_{s,4}(\*r)\;   {y}_{s,5}(\*r)\;  {y}_{s,6}(\*r)  \;  {y}_{s,7}(\*r)\;  {y}_{s,8}(\*r) \;  {y}_{s,9}(\*r)\;  {y}_{s,10}(\*r)\;    {y}_{t,1}(\*r)\; \\ {y}_{ t,2}(\*r)\;  {y}_{t,3}(\*r)\;   {y}_{t,4}(\*r)\;   {y}_{t,5}(\*r)\;  {y}_{t,6}(\*r)\;  {y}_{t,7}(\*r)\;  {y}_{t,8}(\*r)\;  {y}_{t,9}(\*r)  \;   {y}_{b,1}(\*r)\;   {y}_{b,2}(\*r)]^\top$. This simplification is achieved by observing  the structure  $\mathbf{y}^{(k)}$ it inherits from $\mathbf{w}^{(k+1)}$ and $\pmb{\beta}^{(k)}$. Now the cost separated along $g$ and $v$ is given by:
 \begin{align*}
 	\nonumber   \bar{L}_1(g)  = &  \frac{1}{2} \big(\|h*g - y_{m}\|_{2}^2 
 	+  \|d_{xx}*g - y_{s,1}\|_{2}^2+  \|d_{yy}*g - y_{s,2}\|_{2}^2 +\\&
 	\|d_{xy}*g - y_{s,3}\|_{2}^2 +	\|d_{xy}*g - y_{s,4}\|_{2}^2+  \|g - y_{s,5}\|_{2}^2+\|g - y_{b,1}\|_{2}^2 \big)
 \end{align*} 
 \begin{align*}
 	\nonumber   \bar{L}_2(v)  =  & \frac{1}{2} \big(\|d_{xx}*v - y_{s,6}\|_{2}^2 
 	+   \|d_{yy}*v - y_{s,7}\|_{2}^2+ \|d_{xy}*v - y_{s,8}\|_{2}^2 +\|d_{xy}*v - y_{s,9}\|_{2}^2 + \\& 
 	\|v - y_{s,10}\|_{2}^2 +\|d_{xx}*v - y_{t,1}\|_{2}^2 
 	+   \|d_{yy}*v - y_{t,2}\|_{2}^2+ \|d_{xy}*v - y_{t,3}\|_{2}^2 + \\& \|d_{xy}*v - y_{t,4}\|_{2}^2
 	\|d_{yt}*v - y_{t,5}\|_{2}^2 +\|d_{yt}*v - y_{t,6}\|_{2}^2+  \|d_{xt}*v - y_{t,7}\|_{2}^2+\\&\|d_{xt}*v - y_{t,8}\|_{2}^2+ \|d_{tt}*v - y_{t,9}\|_{2}^2 +\|v - y_{b,2}\|_{2}^2 \big) 
 \end{align*}

 For notational convenience, we let $\mathbf{y} = \mathbf{y}^{(k)}$ and $\hat{\mathbf{f}}=\mathbf{f}^{(k+1)}$.
 Recall that $\mathbf{f(\*r)}=[g(\*r) \;\;v(\*r)]^\top$. The function $ \bar{L}_1(g)$ and $ \bar{L}_2(v)$ are quadratic in nature in the variables $g$ and $v$ respectively. The minima of both these functions can be obtained by solving the equations$\nabla_{g}\bar{L}_1(g)=\mathbf{0}$ and $\nabla_{v}\bar{L}_2(v)=\mathbf{0}$ respectively. This requires evaluation of the gradient expressions which are given below.
 
 \begin{align}
 	\nabla_{g}\bar{L}_1(g) = & \tilde{h} *h*g +\tilde{d}_{xx}*d_{xx}*g+\tilde{d}_{yy}*d_{yy}*g+2\tilde{d}_{xy}*d_{xy}*g+g\\\nonumber &-\tilde{h}*y_m	- \tilde{d}_{xx}*y_{s,1}   -\tilde{d}_{yy}*y_{s,2}-\tilde{d}_{xy}*y_{s,3}-\tilde{d}_{xy}*y_{s,4}-y_{s,5}-y_{b,1}
 \end{align}
 \begin{align}
 	\nabla_{v}\bar{L}_2(v) = & \tilde{d}_{xx}*d_{xx}*v+ \tilde{d}_{yy}*d_{yy}*v+2\tilde{d}_{xy}*d_{xy}*v+2v
 	+\tilde{d}_{xt}*d_{xt}*v\\ \nonumber&+\tilde{d}_{yt}*d_{yt}*v+\tilde{d}_{tt}*d_{tt}*v 
 	- \tilde{d}_{xx}*y_{s,5} -\tilde{d}_{yy}*y_{s,6}-\tilde{d}_{xy}*y_{s,7}-\tilde{h}*y_m-y_{s,8} \\ \nonumber&
 	- \tilde{d}_{xt}*y_{t,1} -\tilde{d}_{yt}*y_{t,2}-\tilde{d}_{tt}*y_{t,3}  
 	- \tilde{d}_{xt}*y_{t,4} -\tilde{d}_{yt}*y_{t,5}-\tilde{d}_{tt}*y_{t,6}
 \end{align}

 This completes the solution of the $\*f$ sub-problem which form part of the   ADMM iterates.
 
 \section{Experiments} 	\label{Section : Experiments}
\begin{figure}[!htbp]
 	\centering
 	\includegraphics[width=.7\textwidth]{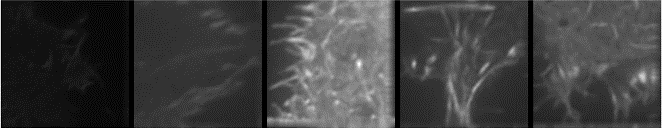}
 	\caption{ The five original   images }
 	\label{Figure:Fiveimages}
 \end{figure}

 To demonstrate the effectiveness of STAIC regularization, we consider restoration of time varying  TIRF images,
 and compare with the method  ICTV, 3D-TV2 and CST. 
 We selected five image sequences obtained from a high Numerical Aperture  (high bandwidth) TIRF microscope under nearly noise-free
 conditions,  and designate them as the ground truth models.  
 The ground truth models are given in   \Cref{Figure:Fiveimages}.
 We then simulate measured images by blurring these models
 with PSF corresponding to low NA systems and by adding mixed Poisson-Gaussian noise as shown below :
 \begin{equation*}
 	m = \mathcal{P}(\gamma_p(h * g))  + \eta 
 \end{equation*} 
 Here  $\gamma_p$ is a  parameter to control the strength of Poisson noise.
 We consider PSF corresponding to  five NA values namely 0.8, 0.9, 1.0, 1.1,  and 1.2. 
 We kept the Gaussian noise level fixed, and considered two $\gamma_p$ values of  1 and 5 to generate the dataset. A higher value of $\gamma_p$ results in a lower level of Poisson noise corruption. This makes a set of 50 measured image sequences generated from our five ground truth images.

 \subsection{Experiment 1}   	 
 
 In first experiment, we choose to demonstrate the advantages gained by treating restoration of time varying images as a separate problem as opposed to treating them under classical image restoration by treating time as the third dimension. To this end we consider 3D-TV2 regularization based restoration of our dataset of 50 images. The resultant restoration scheme which we call 3D-TV2 is posed as the following optimization problem
 \begin{equation}
 	\hat{g}= \argmin{g}{\frac{1}{2}\sum_{i=1}^{n_F}\|(h*g_i)-m_i\|_{F}^2 +\lambda  \sum_{\*r}\|(M*g)(\*r) \|_{F}}
 \end{equation}
 
 where $M(\*r)$ is the 3D Hessian operator defined earlier and $\lambda$ is the regularization parameter. The algorithm is tuned for $\lambda$ to obtain the best SNR. The results are presented in \Cref{Table_M1,Table_M2}   for the two Poisson noise levels $\gamma_p=1,5$ respectively  along with results of  the STAIC scheme. The resultant SNR (in dB) was not found to be   competitive in comparison to our STAIC scheme or even other algorithms we used in  subsequent experiments. The SSIM scores also demonstrate   the shortcomings of this approach.  It  also demonstrates that we indeed need to treat the time dimension differently by considering temporal variations as a different phenomenon in relation to spatial variations in the TIRF image.
 
 \subsection{Experiment 2}
 
 In the second set of experiments we compare the performance of proposed STAIC scheme against the Combined Spatio-Temporal Regularization (CST) and Infimal Convolution TV (ICTV-2DT). In CST   the measured image is subject to both spatial and temporal regularization by employing a sum of norms regularization. The corresponding image restoration optimization problems using CST regularization  takes the form:
 
 \begin{equation}
 	\argmin{g}{ \frac{1}{2}\sum_{i=1}^{n_F}\|(h*g_i)-m_i\|_{F}^2 +\lambda R_{CST}(g)}
 \end{equation}
 In addition, we also generated the restoration results of applying ICTV-2DT regularization scheme. As discussed earlier ICTV belongs to the family of infimal convolution regularization where spatial and temporal components of signal are modeled  differently by the regularization. The ICTV-2DT optimization problem takes the form:
 \begin{equation}
 	\argmin{g}{ \frac{1}{2}\sum_{i=1}^{n_F}\|(h*g_i)-m_i\|_{F}^2 +\lambda R_{ICTV}(g)}
 \end{equation}
 The results of ICTV-2DT, CST and STAIC are aslo presented in \Cref{Table_M1,Table_M2} for the two Poisson noise levels. Our proposed algorithm STAIC perform better than  both CST and ICTV-2DT regularized approaches to restoration for the TIRF 2DT dataset.   STAIC was found to result in a higher restoration quality as measured using  SNR   as well as   SSIM which is another popular  image quality  measure.. It may be noted that SSIM of restored spatio-temporal image  presented in both the tables is the average of SSIM values across the image frames.  Our STAIC achieves SNR improvement over the second best approach in the range of $0.5$ dB to $2$ dB in the 50 image dataset considered. This improvement can be attributed to the improved regularization design that captures the prior of temporal component of the signal in comparison to the other algorithms. In SSIM measure too we have a clear gain over the other approaches considered in our experiments.

 \begin{table}[!htbp]
 	\centering
 	
 	\begin{tabular}{|l  |l |l |l |l |l |l |l |l |l|} \hline  
 		\multicolumn{2}{|l|}{} & \multicolumn{2}{|c|}{STAIC} & \multicolumn{2}{|c|}{ICTV} & \multicolumn{2}{|c|}{CST} & \multicolumn{2}{|c|}{3DTV2} \\ \hline  
 		Image & NA & ssim & snr & ssim & snr & ssim & snr & ssim & snr \\ \hline  
 		\multirow{5}{*}{5135} & 0.8 & 0.853 & 11.58 & 0.758 & 8.84 & 0.841 & 8.80 & 0.656 & 8.48 \\  \cline{2-10} 
 		& 0.9 & 0.873 & 13.01 & 0.781 & 9.91 & 0.863 & 9.86 & 0.669 & 9.45 \\ \cline{2-10}  
 		& 1 & 0.887 & 14.31 & 0.799 & 10.84 & 0.880 & 10.80 & 0.677 & 10.28 \\ \cline{2-10}  
 		& 1.1 & 0.897 & 15.46 & 0.812 & 11.65 & 0.891 & 11.61 & 0.683 & 10.98 \\ \cline{2-10} 
 		& 1.2 & 0.905 & 16.54 & 0.822 & 12.39 & 0.901 & 12.34 & 0.688 & 11.60 \\ \hline  
 		\multirow{5}{*}{5142} & 0.8 & 0.912 & 11.88 & 0.814 & 10.17 & 0.886 & 10.18 & 0.620 & 9.96 \\ \cline{2-10}
 		& 0.9 & 0.926 & 13.26 & 0.831 & 11.30 & 0.905 & 11.31 & 0.629 & 11.01 \\ \cline{2-10} 
 		& 1 & 0.934 & 14.52 & 0.848 & 12.31 & 0.918 & 12.32 & 0.636 & 11.94 \\ \cline{2-10}  
 		& 1.1 & 0.942 & 15.68 & 0.865 & 13.21 & 0.928 & 13.22 & 0.640 & 12.75 \\ \cline{2-10} 
 		& 1.2 & 0.946 & 16.78 & 0.876 & 14.04 & 0.936 & 14.05 & 0.643 & 13.49 \\ \hline  
 		\multirow{5}{*}{5147} & 0.8 & 0.908 & 12.48 & 0.901 & 11.94 & 0.903 & 11.95 & 0.746 & 11.70 \\ \cline{2-10}  
 		& 0.9 & 0.926 & 13.67 & 0.914 & 13.05 & 0.921 & 13.06 & 0.755 & 12.73 \\ \cline{2-10}  
 		& 1 & 0.939 & 14.73 & 0.914 & 14.04 & 0.934 & 14.05 & 0.761 & 13.64 \\ \cline{2-10}  
 		& 1.1 & 0.947 & 15.70 & 0.923 & 14.93 & 0.943 & 14.94 & 0.766 & 14.44 \\ \cline{2-10}  
 		& 1.2 & 0.954 & 16.58 & 0.930 & 15.74 & 0.950 & 15.75 & 0.769 & 15.15 \\ \hline  
 		\multirow{5}{*}{5157} & 0.8 & 0.874 & 8.04 & 0.794 & 7.28 & 0.814 & 7.29 & 0.615 & 6.94 \\ \cline{2-10}  
 		& 0.9 & 0.901 & 9.46 & 0.835 & 8.59 & 0.850 & 8.60 & 0.628 & 8.11 \\ \cline{2-10} 
 		& 1 & 0.919 & 10.75 & 0.854 & 9.76 & 0.877 & 9.76 & 0.638 & 9.11 \\ \cline{2-10}  
 		& 1.1 & 0.932 & 11.90 & 0.844 & 10.81 & 0.897 & 10.79 & 0.644 & 9.96 \\ \cline{2-10}  
 		& 1.2 & 0.941 & 12.96 & 0.853 & 11.75 & 0.912 & 11.73 & 0.649 & 10.70 \\ \hline  
 		\multirow{5}{*}{5158} & 0.8 & 0.943 & 14.38 & 0.918 & 13.18 & 0.927 & 13.14 & 0.513 & 12.00 \\ \cline{2-10}  
 		& 0.9 & 0.954 & 15.76 & 0.917 & 14.37 & 0.940 & 14.33 & 0.519 & 12.86 \\ \cline{2-10} 
 		& 1 & 0.962 & 17.02 & 0.928 & 15.44 & 0.949 & 15.39 & 0.522 & 13.58 \\ \cline{2-10} 
 		& 1.1 & 0.967 & 18.15 & 0.936 & 16.38 & 0.955 & 16.33 & 0.524 & 14.16 \\ \cline{2-10} 
 		& 1.2 & 0.970 & 19.21 & 0.941 & 17.24 & 0.960 & 17.18 & 0.526 & 14.64 \\ \hline 
 		
 	\end{tabular}
 	\caption{The Restoration Quality for Noise Level 1 ($\gamma_p=1$)}
 	\label{Table_M1}
 \end{table}

 \begin{table}[!htbp]
 	\centering
 	
 	\begin{tabular}{|l |l |l |l |l |l |l |l |l |l|} \hline  
 		\multicolumn{2}{|c|}{}& \multicolumn{2}{|c|}{STAIC} & \multicolumn{2}{|c|}{ICTV} & \multicolumn{2}{|c|}{CST} & \multicolumn{2}{|c|}{3DTV2} \\ \hline 
 		Image & NA & ssim & snr& ssim & snr& ssim & snr& ssim & snr\\ \hline 
 		\multirow{5}{*}{5135} & 0.8 & 0.867 & 11.60 & 0.761 & 8.85 & 0.857 & 8.81 & 0.658 & 8.48 \\ \cline{2-10}
 		& 0.9 & 0.889 & 13.03 & 0.789 & 9.91 & 0.881 & 9.87 & 0.671 & 9.45 \\ \cline{2-10} 
 		& 1 & 0.903 & 14.34 & 0.809 & 10.85 & 0.898 & 10.81 & 0.678 & 10.28 \\ \cline{2-10} 
 		& 1.1 & 0.914 & 15.52 & 0.823 & 11.68 & 0.910 & 11.63 & 0.684 & 10.99 \\ \cline{2-10}
 		& 1.2 & 0.922 & 16.59 & 0.835 & 12.41 & 0.920 & 12.36 & 0.689 & 11.60 \\ \hline 
 		\multirow{5}{*}{5142} & 0.8 & 0.929 & 11.89 & 0.892 & 10.19 & 0.892 & 10.18 & 0.621 & 9.96 \\ \cline{2-10}
 		& 0.9 & 0.943 & 13.28 & 0.908 & 11.32 & 0.911 & 11.31 & 0.629 & 11.01 \\ \cline{2-10} 
 		& 1 & 0.953 & 14.55 & 0.920 & 12.33 & 0.925 & 12.32 & 0.636 & 11.94 \\ \cline{2-10} 
 		& 1.1 & 0.960 & 15.72 & 0.929 & 13.24 & 0.936 & 13.22 & 0.640 & 12.75 \\ \cline{2-10}
 		& 1.2 & 0.966 & 16.84 & 0.935 & 14.08 & 0.944 & 14.06 & 0.643 & 13.49 \\ \hline 
 		\multirow{5}{*}{5147} & 0.8 & 0.914 & 12.49 & 0.910 & 11.95 & 0.907 & 11.95 & 0.746 & 11.70 \\\cline{2-10}
 		& 0.9 & 0.932 & 13.68 & 0.924 & 13.07 & 0.925 & 13.06 & 0.756 & 12.74 \\ \cline{2-10}
 		& 1 & 0.945 & 14.75 & 0.935 & 14.06 & 0.938 & 14.06 & 0.762 & 13.64 \\ \cline{2-10} 
 		& 1.1 & 0.954 & 15.71 & 0.942 & 14.95 & 0.948 & 14.94 & 0.766 & 14.44 \\ \cline{2-10} 
 		& 1.2 & 0.961 & 16.60 & 0.948 & 15.77 & 0.955 & 15.76 & 0.770 & 15.15 \\ \hline 
 		\multirow{5}{*}{5157} & 0.8 & 0.881 & 8.04 & 0.795 & 7.28 & 0.815 & 7.29 & 0.615 & 6.94 \\ \cline{2-10} 
 		& 0.9 & 0.908 & 9.47 & 0.836 & 8.59 & 0.852 & 8.60 & 0.628 & 8.11 \\ \cline{2-10}
 		& 1 & 0.927 & 10.76 & 0.855 & 9.77 & 0.879 & 9.77 & 0.638 & 9.12 \\ \cline{2-10} 
 		& 1.1 & 0.940 & 11.91 & 0.869 & 10.81 & 0.899 & 10.80 & 0.644 & 9.96 \\ \cline{2-10}
 		& 1.2 & 0.949 & 12.97 & 0.854 & 11.76 & 0.914 & 11.73 & 0.649 & 10.70 \\ \hline 
 		\multirow{5}{*}{5158} & 0.8 & 0.949 & 14.39 & 0.927 & 13.20 & 0.931 & 13.14 & 0.514 & 12.00 \\ \cline{2-10}
 		& 0.9 & 0.961 & 15.78 & 0.939 & 14.40 & 0.945 & 14.34 & 0.519 & 12.86 \\ \cline{2-10}
 		& 1 & 0.969 & 17.04 & 0.947 & 15.48 & 0.954 & 15.40 & 0.522 & 13.57 \\ \cline{2-10}
 		& 1.1 & 0.974 & 18.19 & 0.953 & 16.44 & 0.960 & 16.35 & 0.525 & 14.16 \\ \cline{2-10} 
 		& 1.2 & 0.978 & 19.26 & 0.958 & 17.30 & 0.965 & 17.20 & 0.526 & 14.64 \\ \hline
 		
 	\end{tabular}
 	\caption{The Restoration Quality for Noise Level 2 $(\gamma_p=5)$}
 	\label{Table_M2}
 \end{table}

 \section{Conclusion}
 We proposed a spatio-temporal regularization method  for restoration of time varying 2D  images. Our design of the regularization terms led to an underlying additive decomposition of the signal enabling differentiated regularization of regions with motion and without motion. The effectiveness of this decomposition is evident in the superior reconstruction quality evaluated by structure similarity as well as signal to noise ratio measures of image quality. Our method was found to perform better than other algorithms each designed using different philosophy towards spatio-temporal signal restoration. This better quality reconstruction was enabled through design of  a fast and efficient algorithm for minimizing the resultant regularized cost  using the ADMM framework.  The proposed algorithms effectiveness can be attributed to the advantages provided by the infimal convolution framework in exploiting correlation across frames inherent to spatio-temporal signals.

 \appendix  
 \section{proof} \label{Appendix : Proofs}
 \begin{proof}   	
 	Let us start by considering the optimization problem   	
 	\begin{equation} 
 		\underset{{\mathbf z}\in \mathbb{R}^{10}}{\operatorname{argmin}} \;\;{L}_{s}(\mathbf{z},\mathbf{y},\alpha_s)
 	\end{equation}
 	where 	${L}_{s}(\mathbf{z}, \mathbf{y}, \alpha_s)= \frac{\rho}{2} \|\mathbf{y}-\mathbf{z}\|_{2}^2+ \sqrt{2} \alpha_s \|A_s\mathbf{z}\|_{{2}}$. Note that $\mathbf{y,z} \in \mathbb{R}^{10}$. 
 	
 	Consider symmetric matrix $A_s^\top A_s$ which is orthogonally diagonalizable. In other words, we can find   an orthogonal matrix of eigenvectors $P$  and  a diagonal matrix of eigenvalues $D$ such that  $A_s^\top A_s=PDP^\top$.
 	
 	Since $A_s^\top  A_s$ can be derived from given matrix $A_s$, we can derive $P$ and $D$ by eigen decomposition and is given below:

 		\begin{equation} \label{Defintion : Matrix  P of eigenvectors}
 			P= \left[ \begin{array}{cccccccccc}
 				\frac{1}{\sqrt{2}} & 0 & 0 & 0 & 0 &-\frac{1}{\sqrt{2}} &0 &0 &0 & 0 \\
 				0 &\frac{1}{\sqrt{2}} & 0 & 0 & 0 & 0 &-\frac{1}{\sqrt{2}} &0 &0 &0   \\
 				0 & 0 &\frac{1}{\sqrt{2}} & 0 & 0 & 0 & 0 &-\frac{1}{\sqrt{2}} &0 &0    \\
 				0&0 & 0 &\frac{1}{\sqrt{2}} & 0 & 0 & 0 & 0 &-\frac{1}{\sqrt{2}} &0    \\
 				0&	  0&0 & 0 &\frac{1}{\sqrt{2}} & 0 & 0 & 0 & 0 &-\frac{1}{\sqrt{2}}     \\
 				-\frac{1}{\sqrt{2}} & 0 & 0 & 0 & 0 &\frac{1}{\sqrt{2}} &0 &0 &0 & 0 \\
 				0 &-\frac{1}{\sqrt{2}} & 0 & 0 & 0 & 0 &\frac{1}{\sqrt{2}} &0 &0 &0   \\
 				0 & 0 &-\frac{1}{\sqrt{2}} & 0 & 0 & 0 & 0 &\frac{1}{\sqrt{2}} &0 &0    \\
 				0&0 & 0 &-\frac{1}{\sqrt{2}} & 0 & 0 & 0 & 0 &\frac{1}{\sqrt{2}} &0    \\
 				0&	  0&0 & 0 &-\frac{1}{\sqrt{2}} & 0 & 0 & 0 & 0 &\frac{1}{\sqrt{2}}     \\
 			\end{array}  \right]\text{ and } D= diag(\left[ \begin{array}{cccccccc}
 				1  \\
 				1  \\
 				1   \\
 				1   \\
 				1   \\
 				0  \\
 				0  \\
 				0   \\
 				0  \\
 				0  
 			\end{array}  \right])
 		\end{equation}
 		
 		where $\operatorname{diag}$ is the operation  that creates a diagonal matrix with its input vector forming the diagonal entries. We define $\hat{\mathbf{z}}=P^\top \mathbf{z}$ and $\hat{\mathbf{y}}=P^\top \mathbf{y}$. Since $P$ is orthogonal, it is also true that $\|P^\top \mathbf{x}\|_2=\|\mathbf{x}\|_2$.
 		Since $\|\mathbf{x}\|_2 =\sqrt{ \langle \mathbf{x},\mathbf{x}\rangle }$ where $\langle \cdot ,\cdot \rangle$ is the standard inner product, it can be seen that 
 		\begin{equation} \|A_s\mathbf{z}\|_2= \sqrt{ \langle A_s\mathbf{z},A_s\mathbf{z} \rangle }  = \sqrt{ \langle \mathbf{z},A_s^\top A_s\mathbf{z} \rangle } = \sqrt{\mathbf{z}^\top A_s^\top A_s\mathbf{z} } = \sqrt{\mathbf{z}^\top PDP^\top \mathbf{z}}  =\sqrt{\hat{\mathbf{z}}^\top D\hat{\mathbf{z}}}\end{equation}
 		
 		Also observe that:
 		
 		\begin{equation}
 			\frac{\rho}{2} \|\mathbf{y}-\mathbf{z}\|_{2}^2 =  \frac{\rho}{2} \|P^\top \mathbf{y}-P^\top \mathbf{z}\|_{2}^2=\frac{\rho}{2} \| \hat{\mathbf{y}}-\hat{\mathbf{z}}\|_{2}^2
 		\end{equation}

 		Let us define $\hat{\mathbf{z}}_1 ,\hat{\mathbf{z}}_2 \in \mathbb{R}^5$ such that $\hat{\mathbf{z}}= \left[\begin{array}{c} \hat{\mathbf{z}}_1 \\ \hat{\mathbf{z}}_2\end{array} \right]$. Also define
 		$\hat{\mathbf{y}_1}  ,\hat{\mathbf{y}_2}  \in \mathbb{R}^4$ such that $\hat{\mathbf{y}}= \left[\begin{array}{c} \hat{\mathbf{y}_1} \\ \hat{\mathbf{y}_2}\end{array} \right]$

 		The original optimization problem in terms of $\hat{\mathbf{z}}$ is given by 
 		
 		\begin{equation} 
 			\underset{{\mathbf z}\in \mathbb{R}^{10}}{\operatorname{argmin}} \;\; \frac{\rho}{2} \| \hat{\mathbf{y}}-\hat{\mathbf{z}}\|_{2}^2 +\sqrt{2} \alpha_s \sqrt{\hat{\mathbf{z}}D\hat{\mathbf{z}}}
 		\end{equation}
 		By appealing to the sub vectors definition in $\hat{\mathbf{z}}$ and $\hat{\mathbf{y}}$, the problem can be stated as 
 		\begin{equation}
 			\underset{\hat{ \mathbf{z}}_1,\hat{ \mathbf{z}}_2 \in \mathbb{R}^5}{\operatorname{argmin}} \;\; \frac{\rho}{2} \| \hat{\mathbf{y}_1} -\hat{\mathbf{z}}_1\|_{2}^2 +\frac{\rho}{2} \| \hat{\mathbf{y}_2} -\hat{\mathbf{z}}_2\|_{2}^2+ \sqrt{2} \alpha_s \|{\hat{\mathbf{z}_1}}\|_2
 		\end{equation}
 		By considering two separate optimization problems in $\hat{\mathbf{z}}_1$ and $\hat{\mathbf{z}}_2$, we get
 		\begin{eqnarray}
 			&\underset{\hat{ \mathbf{z}}_1\in \mathbb{R}^5}{\operatorname{argmin}} \;\; \frac{\rho}{2} \| \hat{\mathbf{y}_1} -\hat{\mathbf{z}}_1\|_{2}^2 + \sqrt{2} \alpha_s \|{\hat{\mathbf{z}_1}}\|_2\\
 			&\underset{\hat{ \mathbf{z}}_2 \in \mathbb{R}^5}{\operatorname{argmin}} \;\; \frac{\rho}{2} \| \hat{\mathbf{y}_2} -\hat{\mathbf{z}}_2\|_{2}^2
 		\end{eqnarray}   		
 		The optimization problem in variable $\hat{\mathbf{z}}_1$ is simply the proximal operator of $\ell_2$ norm which is well known \cite{parikh2014proximal} and is given by
 		\begin{equation}
 			\hat{z}_1^* =\gamma \hat{\mathbf{y}_1}
 		\end{equation}
 		where $\gamma =  \max(0,1 -\frac{\sqrt{2} \alpha_s}{\rho \|\hat{\mathbf{y}_1}\|_2})$. 
 		The solution of  optimization problem in variable $\hat{\mathbf{z}}_2$ is given by $	\hat{\mathbf{z}}_2^*=\hat{\mathbf{y}_2}$. 
 		Now, we map back to space of variable $\mathbf{z}$ by multiplying with $P$ \begin{equation}
 			{\mathbf{z}}^* = P	\hat{\mathbf{z}}^*
 		\end{equation}
 		where $	\hat{z}^*=\left[\begin{array}{c} \gamma \hat{\mathbf{y}_1} \\ \hat{\mathbf{y}_2}\end{array} \right]$.
 		
 	\end{proof}

\bibliographystyle{unsrtnat}
\bibliography{ citation}  

\begin{thebibliography}{26}
\providecommand{\natexlab}[1]{#1}
\providecommand{\url}[1]{\texttt{#1}}
\expandafter\ifx\csname urlstyle\endcsname\relax
  \providecommand{\doi}[1]{doi: #1}\else
  \providecommand{\doi}{doi: \begingroup \urlstyle{rm}\Url}\fi

\bibitem[Katsaggelos(1989)]{katsaggelos1989iterative}
Aggelos~K Katsaggelos.
\newblock Iterative image restoration algorithms.
\newblock \emph{Optical engineering}, 28\penalty0 (7):\penalty0 735--748, 1989.

\bibitem[Engl et~al.(1996)Engl, Hanke, and Neubauer]{engl1996regularization}
Heinz~Werner Engl, Martin Hanke, and Andreas Neubauer.
\newblock \emph{Regularization of inverse problems}, volume 375.
\newblock Springer Science \& Business Media, 1996.

\bibitem[Rangayyan(2004)]{rangayyan2004biomedical}
Rangaraj~M Rangayyan.
\newblock \emph{Biomedical image analysis}.
\newblock CRC press, 2004.

\bibitem[Ramani et~al.(2012)Ramani, Liu, Rosen, Nielsen, and
  Fessler]{ramani2012regularization}
Sathish Ramani, Zhihao Liu, Jeffrey Rosen, Jon-Fredrik Nielsen, and Jeffrey~A
  Fessler.
\newblock Regularization parameter selection for nonlinear iterative image
  restoration and mri reconstruction using gcv and sure-based methods.
\newblock \emph{IEEE Transactions on Image Processing}, 21\penalty0
  (8):\penalty0 3659--3672, 2012.

\bibitem[Viswanath et~al.(2020)Viswanath, Ghulyani, De~Beco, Dahan, and
  Arigovindan]{viswanath2020image}
Sanjay Viswanath, Manu Ghulyani, Simon De~Beco, Maxime Dahan, and Muthuvel
  Arigovindan.
\newblock Image restoration by combined order regularization with optimal
  spatial adaptation.
\newblock \emph{IEEE Transactions on Image Processing}, 29:\penalty0
  6315--6329, 2020.

\bibitem[Arigovindan(2013)]{arigovindan2013image}
Muthuvel Arigovindan.
\newblock Image deconvolution research: its scope and importance in live cell
  microscopy.
\newblock \emph{Current Science}, pages 1501--1511, 2013.

\bibitem[Li et~al.(2018)Li, Xue, and Blu]{li2018accurate}
Jizhou Li, Feng Xue, and Thierry Blu.
\newblock Accurate 3d psf estimation from a wide-field microscopy image.
\newblock In \emph{2018 IEEE 15th International Symposium on Biomedical Imaging
  (ISBI 2018)}, pages 501--504. IEEE, 2018.

\bibitem[Fan et~al.(2019)Fan, Huang, Li, Chen, and Tan]{fan2019one}
Junchao Fan, Xiaoshuai Huang, Liuju Li, Liangyi Chen, and Shan Tan.
\newblock One-step deconvolution for multi-angle tirf microscopy with enhanced
  resolution.
\newblock \emph{Biomedical optics express}, 10\penalty0 (3):\penalty0
  1097--1110, 2019.

\bibitem[Lefkimmiatis et~al.(2011)Lefkimmiatis, Bourquard, and
  Unser]{lefkimmiatis2011hessian}
Stamatios Lefkimmiatis, Aur{\'e}lien Bourquard, and Michael Unser.
\newblock Hessian-based norm regularization for image restoration with
  biomedical applications.
\newblock \emph{IEEE Transactions on Image Processing}, 21\penalty0
  (3):\penalty0 983--995, 2011.

\bibitem[Lefkimmiatis et~al.(2013)Lefkimmiatis, Ward, and
  Unser]{lefkimmiatis2013hessian}
Stamatios Lefkimmiatis, John~Paul Ward, and Michael Unser.
\newblock Hessian schatten-norm regularization for linear inverse problems.
\newblock \emph{IEEE transactions on image processing}, 22\penalty0
  (5):\penalty0 1873--1888, 2013.

\bibitem[Lindsten et~al.(2011)Lindsten, Ohlsson, and
  Ljung]{lindsten2011clustering}
Fredrik Lindsten, Henrik Ohlsson, and Lennart Ljung.
\newblock Clustering using sum-of-norms regularization: With application to
  particle filter output computation.
\newblock In \emph{2011 IEEE Statistical Signal Processing Workshop (SSP)},
  pages 201--204. IEEE, 2011.

\bibitem[Holler and Kunisch(2014)]{holler2014infimal}
Martin Holler and Karl Kunisch.
\newblock On infimal convolution of tv-type functionals and applications to
  video and image reconstruction.
\newblock \emph{SIAM Journal on Imaging Sciences}, 7\penalty0 (4):\penalty0
  2258--2300, 2014.

\bibitem[Guo et~al.(2014)Guo, Qin, and Yin]{guo2014new}
Weihong Guo, Jing Qin, and Wotao Yin.
\newblock A new detail-preserving regularization scheme.
\newblock \emph{SIAM journal on imaging sciences}, 7\penalty0 (2):\penalty0
  1309--1334, 2014.

\bibitem[Bredies et~al.(2010)Bredies, Kunisch, and Pock]{bredies2010total}
Kristian Bredies, Karl Kunisch, and Thomas Pock.
\newblock Total generalized variation.
\newblock \emph{SIAM Journal on Imaging Sciences}, 3\penalty0 (3):\penalty0
  492--526, 2010.

\bibitem[Ulyanov et~al.(2018)Ulyanov, Vedaldi, and Lempitsky]{ulyanov2018deep}
Dmitry Ulyanov, Andrea Vedaldi, and Victor Lempitsky.
\newblock Deep image prior.
\newblock In \emph{Proceedings of the IEEE conference on computer vision and
  pattern recognition}, pages 9446--9454, 2018.

\bibitem[Xing et~al.(2017)Xing, Xie, Su, Liu, and Yang]{xing2017deep}
Fuyong Xing, Yuanpu Xie, Hai Su, Fujun Liu, and Lin Yang.
\newblock Deep learning in microscopy image analysis: A survey.
\newblock \emph{IEEE transactions on neural networks and learning systems},
  29\penalty0 (10):\penalty0 4550--4568, 2017.

\bibitem[von Chamier et~al.(2021)von Chamier, Laine, Jukkala, Spahn, Krentzel,
  Nehme, Lerche, Hern{\'a}ndez-P{\'e}rez, Mattila, Karinou,
  et~al.]{von2021democratising}
Lucas von Chamier, Romain~F Laine, Johanna Jukkala, Christoph Spahn, Daniel
  Krentzel, Elias Nehme, Martina Lerche, Sara Hern{\'a}ndez-P{\'e}rez, Pieta~K
  Mattila, Eleni Karinou, et~al.
\newblock Democratising deep learning for microscopy with zerocostdl4mic.
\newblock \emph{Nature communications}, 12\penalty0 (1):\penalty0 1--18, 2021.

\bibitem[Liu et~al.(2021)Liu, Jin, Chen, Fang, Ablameyko, Yin, and
  Xu]{liu2021survey}
Zhichao Liu, Luhong Jin, Jincheng Chen, Qiuyu Fang, Sergey Ablameyko, Zhaozheng
  Yin, and Yingke Xu.
\newblock A survey on applications of deep learning in microscopy image
  analysis.
\newblock \emph{Computers in Biology and Medicine}, 134:\penalty0 104523, 2021.

\bibitem[Hoffman et~al.(2021)Hoffman, Slavitt, and
  Fitzpatrick]{hoffman2021promise}
David~P Hoffman, Isaac Slavitt, and Casey~A Fitzpatrick.
\newblock The promise and peril of deep learning in microscopy.
\newblock \emph{Nature methods}, 18\penalty0 (2):\penalty0 131--132, 2021.

\bibitem[Schloegl et~al.(2017)Schloegl, Holler, Schwarzl, Bredies, and
  Stollberger]{schloegl2017infimal}
Matthias Schloegl, Martin Holler, Andreas Schwarzl, Kristian Bredies, and
  Rudolf Stollberger.
\newblock Infimal convolution of total generalized variation functionals for
  dynamic mri.
\newblock \emph{Magnetic resonance in medicine}, 78\penalty0 (1):\penalty0
  142--155, 2017.

\bibitem[Zhang et~al.(2018)Zhang, Li, Krol, Schmidtlein, Lipson, Feiglin, and
  Xu]{zhang2018infimal}
Jiahan Zhang, Si~Li, Andrzej Krol, C~Ross Schmidtlein, Edward Lipson, David
  Feiglin, and Yuesheng Xu.
\newblock Infimal convolution-based regularization for spect reconstruction.
\newblock \emph{Medical physics}, 45\penalty0 (12):\penalty0 5397--5410, 2018.

\bibitem[Tr{\'e}moulh{\'e}ac et~al.(2014)Tr{\'e}moulh{\'e}ac, Dikaios,
  Atkinson, and Arridge]{tremoulheac2014dynamic}
Benjamin Tr{\'e}moulh{\'e}ac, Nikolaos Dikaios, David Atkinson, and Simon~R
  Arridge.
\newblock Dynamic mr image reconstruction--separation from undersampled
  (k,t)-space via low-rank plus sparse prior.
\newblock \emph{IEEE transactions on medical imaging}, 33\penalty0
  (8):\penalty0 1689--1701, 2014.

\bibitem[Chen et~al.(2022)Chen, Wang, Chen, Chen, and Sun]{chen2022pyramid}
Eric~Z Chen, Puyang Wang, Xiao Chen, Terrence Chen, and Shanhui Sun.
\newblock Pyramid convolutional rnn for mri image reconstruction.
\newblock \emph{IEEE Transactions on Medical Imaging}, 2022.

\bibitem[Arigovindan et~al.(2013)Arigovindan, Fung, Elnatan, Mennella, Chan,
  Pollard, Branlund, Sedat, and Agard]{arigovindan2013high}
Muthuvel Arigovindan, Jennifer~C Fung, Daniel Elnatan, Vito Mennella,
  Yee-Hung~Mark Chan, Michael Pollard, Eric Branlund, John~W Sedat, and David~A
  Agard.
\newblock High-resolution restoration of 3d structures from widefield images
  with extreme low signal-to-noise-ratio.
\newblock \emph{Proceedings of the National Academy of Sciences}, 110\penalty0
  (43):\penalty0 17344--17349, 2013.

\bibitem[Boyd et~al.(2011)Boyd, Parikh, and Chu]{boyd2011distributed}
Stephen Boyd, Neal Parikh, and Eric Chu.
\newblock \emph{Distributed optimization and statistical learning via the
  alternating direction method of multipliers}.
\newblock Now Publishers Inc, 2011.

\bibitem[Parikh et~al.(2014)Parikh, Boyd, et~al.]{parikh2014proximal}
Neal Parikh, Stephen Boyd, et~al.
\newblock Proximal algorithms.
\newblock \emph{Foundations and trends in Optimization}, 1\penalty0
  (3):\penalty0 127--239, 2014.

\end{thebibliography}

\end{document}